\documentclass[twocolumn,aps,prl,showpacs,superscriptaddress]{revtex4}
\usepackage[LGR,T1]{fontenc}
\usepackage[latin9]{inputenc}
\usepackage{amsmath}
\usepackage{amssymb}
\usepackage{graphicx}
\usepackage{epstopdf}
\makeatletter
\@ifundefined{textcolor}{}
{%
 \definecolor{BLACK}{gray}{0}
 \definecolor{WHITE}{gray}{1}
 \definecolor{RED}{rgb}{1,0,0}
 \definecolor{GREEN}{rgb}{0,1,0}
 \definecolor{BLUE}{rgb}{0,0,1}
 \definecolor{CYAN}{cmyk}{1,0,0,0}
 \definecolor{MAGENTA}{cmyk}{0,1,0,0}
 \definecolor{YELLOW}{cmyk}{0,0,1,0}
}

\usepackage{dcolumn}
\usepackage{bm}

\makeatother

\begin{document}

\title{Kondo effect goes anisotropic in vanadate oxide superlattices}

\author{H.~Rotella}

\altaffiliation[Present address: ]{LETI-CEA, 17 rue des Martyrs, 38054 Grenoble cedex 9, France.}

\affiliation{Laboratoire CRISMAT, CNRS UMR 6508, ENSICAEN et Universit$\acute{e}$
de Caen, 6 Bd Mar$\acute{e}$chal Juin, 14050 Caen Cedex 4, France.}

\author{A.~Pautrat}

\email{alain.pautrat@ensicaen.fr}

\thanks{Corresponding author}

\affiliation{Laboratoire CRISMAT, CNRS UMR 6508, ENSICAEN et Universit$\acute{e}$
de Caen, 6 Bd Mar$\acute{e}$chal Juin, 14050 Caen Cedex 4, France.}

\author{O.~Copie}

\altaffiliation[Present address: ]{CEA, DSM/IRAMIS/SPEC, F-91191 Gif-sur-Yvette Cedex, France }

\affiliation{Laboratoire CRISMAT, CNRS UMR 6508, ENSICAEN et Universit$\acute{e}$
de Caen, 6 Bd Mar$\acute{e}$chal Juin, 14050 Caen Cedex 4, France.}

\author{P.~Boullay}

\affiliation{Laboratoire CRISMAT, CNRS UMR 6508, ENSICAEN et Universit$\acute{e}$
de Caen, 6 Bd Mar$\acute{e}$chal Juin, 14050 Caen Cedex 4, France.}

\author{A.~David}

\affiliation{Laboratoire CRISMAT, CNRS UMR 6508, ENSICAEN et Universit$\acute{e}$
de Caen, 6 Bd Mar$\acute{e}$chal Juin, 14050 Caen Cedex 4, France.}

\author{B.~Mercey}

\affiliation{Laboratoire CRISMAT, CNRS UMR 6508, ENSICAEN et Universit$\acute{e}$
de Caen, 6 Bd Mar$\acute{e}$chal Juin, 14050 Caen Cedex 4, France.}

\author{M.~Morales}

\affiliation{Laboratoire CIMAP, CNRS UMR 6252, ENSICAEN et Universit$\acute{e}$
de Caen, 6 Bd Mar$\acute{e}$chal Juin, 14050 Caen Cedex 4, France.}

\author{W.~Prellier}

\affiliation{Laboratoire CRISMAT, CNRS UMR 6508, ENSICAEN et Universit$\acute{e}$
de Caen, 6 Bd Mar$\acute{e}$chal Juin, 14050 Caen Cedex 4, France.}

\begin{abstract}
We study the transport properties in SrVO$_{3}$/LaVO$_{3}$ (SVO/LVO)
superlattices deposited on SrTiO$_{3}$ (STO) substrates. We show that the electronic conduction occurs in the metallic LVO layers
 with a galvanomagnetism typical of a 2D  Fermi surface.
In addition, a Kondo-like component appears in both the thermal variation of resistivity and the magnetoresistance. Surprisingly, in this system where the STO interface does not contribute to the measured conduction, the Kondo correction is strongly anisotropic. We show that the growth temperature allows a direct control of this contribution. Finally, the key role of vanadium mixed valency stabilized by oxygen vacancies is enlightened.
\end{abstract}

\pacs{PACS numbers: 73.21.Cd,73.50.Jt,72.15.Qm}

\maketitle

In the past decade, there has been a strong interest on the
electronic and magnetic properties that can emerge at oxides interface. In LaAlO$_{3}$/SrTiO$_{3}$ (LAO/STO),
under extremely controlled growth conditions, a conducting sheet with large electronic
mobility ($\mu$ $\sim$ 10$^{3}$-10$^{4}$cm$^{2}$/V.s) is observed between these insulating materials, with typical characteristics of a 2D
 electron gas (2DEG) and a superconducting phase is evidenced at very low temperature. Most of the literature
agrees with the scenario which involved carriers arising
from an electronic reconstruction, a feature needed to avoid the polar discontinuity between the two
perovskite blocks \cite{ohtomo04,nakagawa06,reyren07,kourkoutis06}.
It was also shown that the n-type doping of the STO by oxygen vacancies needs to be carefully
considered to interpret the results \cite{kalabukov}. Among the various features occurring at an oxide interface, unexpected magnetism was reported whereas the two parent compounds are non-magnetic. Long-range ferromagnetism has been inferred from macroscopic measurements up to room temperature \cite{ariando1}.  Contrarily, only micronic and dilute magnetic puddles were seen at low temperature using a local magnetometry \cite{bert}. Peculiar magneto-transport properties were also observed:  A Kondo-like scattering at T$<$50 K, that involves the existence of dilute magnetic scattering centers, and a magnetic hysteresis in the sheet resistance at T$<$0.3 K, attributed to ordered spins or magnetic domains \cite{brinkman}. 
A negative magnetoresistance with extreme anisotropy has been also measured. So far, it has been attributed to the emergence of magnetic states in the 2DEG \cite{shalom}. Besides, experimental evidence suggests that magnetism can be quenched by annealing confirming the important role of oxygen vacancies \cite{annealing}.
The strong electronic mobility in all-oxide structures is considered to be a specific feature of the STO compound, and the interfacial magnetism is attributed to extra electrons in the Ti band with a significant role assigned to ill-defined defects \cite{coey}. However, it is not clear if signatures of similar interfacial magnetism can be observed in a system where STO has not a major role, e.g with magnetic ions other than Ti 3\emph{d}.  

Among the various oxides, vanadate perovskites are interesting systems owing to the possible spin-charge-orbital
coupling. LaVO$_{3}$ (LVO) is an antiferromagnetic Mott insulator with V$^{3+}$ in a 3\emph{d}$^{2}$ electronic configuration
whereas SrVO$_{3}$ (SVO) is a Pauli metal with V$^{4+}$ in a \emph{d}$^{1}$ electronic configuration. Interestingly,
the La$_{1-x}$Sr$_{x}$VO$_{3}$ solid solution exists over the entire
composition (0$\leq$x$\leq$1) without disrupting the distorted perovskite structural network. Also, La$_{1-x}$Sr$_{x}$VO$_{3}$
was widely studied for its metal-insulator-transition (MIT) which can be tuned by carriers doping
\cite{inaba95,miyasaka00}. Substitution of La$^{3+}$
by Sr$^{2+}$ introduces a mixed V$^{3+}$/V$^{4+}$ valency which
favors a metallic behavior \cite{sayer75}.
Another approach for controlling the electronic properties of this system is the preparation of superlattices (SLs), where the cations are artificially ordered in the layers.
Synthesis and characterization of (LVO)$_{m}$/(SVO)$_{n}$ SLs has indeed been previously reported using the pulsed laser deposition (PLD) technique \cite{sheets07,boullay11}, and high quality interfaces were obtained as seen from the Laue fringe in x-rays diffraction and electron microscopy. The question of valency changes close to the interface is more delicate, but we have recently obtained a quantitative map of the oxidation state of the vanadium in (LaVO$_{3}$)$_{6}$/(SrVO$_{3}$)$_{3}$ SLs. The interfaces are sharp in terms of composition, but a mixed valency V$^{3+}$/V$^{4+}$ is extended over three unit cells showing that charge transfert is consistent with an electronic
reconstruction \cite{tan13}. However, the consequences on the
magnetotransport properties have not been yet investigated .

Here, we report a detailed investigation of magnetotransport properties
of a series of (LaVO$_{3}$)$_{6}$/(SrVO$_{3}$)$_{6}$ SLs . Low temperature electronic properties present
a strong anisotropic Kondo correction, similarly to the LAO/STO system, which occurs without any
2D electron gas characteristics due to the the preferential conduction in the SVO layers.
Remarkably, the strenght of the Kondo effect can be directly tuned by the growth conditions in a way suggesting that oxygen vacancies (i) have
a key role, and (ii) stabilize local valency fluctuations of magnetic ions.

Epitaxial (LaVO$_{3}$)$_{6}$/(SrVO$_{3}$)$_{6}$ superlattices were prepared by the Pulsed Laser Deposition (PLD) technique on (001)-oriented
SrTiO$_{3}$ (STO) substrates (cubic with a=3.905 \AA ). Briefly, a KrF laser ($\lambda$=248 nm) with a repetition rate of 3 Hz was
focused alternatively onto a LaVO$_{4}$ and a Sr$_{2}$V$_{2}$O$_{7}$
polycrystalline targets at a fluence of $\simeq$ 2 J/cm$^{2}$. The
substrate is kept at a temperature ranging from 550-750$^{\circ}{C}$
under a dynamic vacuum around 10$^{-5}$ mbar. The target-substrate
distance is fixed at 8.5 cm. The deposition rates are 0.25 \AA /pulse and 0.46 \AA /pulse, for LaVO$_{3}$
(LVO) and for SrVO$_{3}$ (SVO), respectively. We grew a series of [(LaVO$_{3}$)$_{6}$/(SrVO$_{3}$)$_{6}$]$_{18}$
 SLs with a total thickness of 80 nm. The X-ray diffraction
(XRD) measurements were performed with a $\theta$/2$\theta$ monochromated
diffractometer Seifert XRD 3000P ($\lambda$=1.5406 \AA ) for the
1D XRD patterns and with a 4-circles diffractometer MRD PANalytical
($\lambda$=1.5406 \AA ) for high resolution reciprocal space mappings.
Measurements and discussion of the structural properties can be found in the supplemental material (I) \cite{supp}.
Transport and galvanomagnetic properties were measured as a function of temperature ($T$) and magnetic field ($B$) using a Quantum Design PPMS. The Van der Pauw configuration was used for resistivity and Hall effects measurement and a four probe geometry was preferred for the angular magnetoresistance measurements. Electric contacts were made by wire-bonding using an Au/Al alloy wire.

Fig.1a shows the temperature dependance of the resistivity
in the range of 2-300 K for all the series of (LVO)$_{6}$/(SVO)$_{6}$ SLs.
 To extract a resistivity $\rho$ value, we assume that the conducting thickness is that of the SVO layers, as detailed and justified hereafter.  Room temperature resistivity values are in 
the 150-225 $\mu\Omega$.cm range with ratio RRR=$\rho(300K)$/$\rho(2K)$ $\approx$ 1.4-1.5.
At high temperature ($T\geq$50 K), all resistivities
vary as $\rho(T)$=$\rho_{0}$+A.$T^{2}$  (Fig.1b), a typical law for Fermi
liquids where the residual resistivity $\rho_{0}$ is attributed to
the (elastic) electron-impurity scattering, and the A coefficient
quantifies the electron-electron interactions via the effective mass
renormalisation. All SLs present a clear metallic behavior, and the most obvious difference between all them is
the significant decrease of $\rho_{0}$ when the growth
temperature increases. This is in agreement with an increase of the SLs quality
when the growth temperature is higher. We also observe that A
is not constant within the series, and varies quasi-linearly with $\rho_{0}$ (see  Fig. S3  in the supplemental  material \cite{supp}). This can not be explained by a pure Fermi Liquid behavior \cite{nunez} and indicates that inelastic scattering against impurity, the so-called Koshino-Taylor effect, needs to be considered to extract the
intrinsic value of A \cite{nunez,alain}. As detailed in the section (II) of the supplemental material \cite{supp}, we finally calculate that A=4.55 10$^{-4}$ $\mu\varOmega$.cm.K$^{-2}$
in very good agreement with the value of crystalline SVO (A $\approx$ 4.2 10$^{-4}$ $\mu\varOmega$.cm.K$^{-2}$). Resistivity values $\rho$ are also fully consistent with bulk SVO \cite{reyes00}. Finally, from Hall effect measurements, and assuming a single carrier analysis, we extract at room temperature a n-type carrier
density $n_{2D}\approx$10$^{17}$/cm$^{2}$  ($n_{3D}\approx$10$^{22}$/cm$^{2}$) and a low electronic mobility
$\mu\approx$2 cm$^{2}$/V.S, values that are also typical of SVO \cite{gu} (see Fig.2).
Despite the existence of charge transfer \cite{tan13}, the electronic mobility
 appears much lower than the values reported in highly conducting STO-based interfaces,  
confirming that the presence of STO is the key point to have
the strongly mobile carriers. We thus conclude the electronic conduction
of the SL is dominated by the SVO contribution. Note that ultrathin
layers of SVO tends to be insulating by bandwidth controlled dimensional
effect \cite{Yoshi} or by orbital symmetry changes \cite{LVOMIT},
but a thickness of 6 layers is still in the metallic limit in agreement
with our observation.

Let us now focus on the low-temperature regime where an increase of the
resistivity is observed for a decreasing temperature. This variation appears linear with the logarithm of the temperature (Fig.3), and can be written as -B$_K$.log(T). This typical dependence can be attributed to a 2D charge
carriers weak localization or to a single-impurity Kondo effect. In a series of sample, the evolution of the B$_K$  coefficient could distinguish between the two mechanisms.
 In the simplest form of weak localization, this coefficient is indeed an universal constant, and in the
Kondo mechanism, B$_K$ is proportional to the concentration of diluted
magnetic impurities. In the dilute and elastic limits, $\rho_0$ is proportional to the concentration of defects and impurities that scatter the carriers. 
This includes the contribution from both ordinary impurities $\rho_i$, and magnetic impurities $\rho_K$. Thus, the total residual resistivity can be written as $\rho_0$=$\rho_i$+$\rho_K$.
 The evolution of the B$_K$ coefficient as a function of residual resistivity $\rho_0$ shown in the inset of Fig.3 is qualitatively in agreement with a Kondo scenario, as discussed hereafter.

The low-temperature (2-30K) evolution of the resistance for the SL sample grown at 750$^{\circ}{C}$ is presented in Fig.4. The data are well fitted using a Fermi Liquid law with an additional Kondo resistance in the dilute limit like the following \cite{costi94}:

\begin{equation}
R(T)=R_{0}+A.T^{2}+R_{K}
\end{equation}

with 
\begin{equation}
R_{K}(T)=R_{K0}\left(\frac{T_{K}'^2}{T^{2}+T_{K}'^2}\right)^{s} \label{equakondo}
\end{equation}
and 
\begin{equation}
T_{K}'=\frac{T_{K}}{(2^{\frac{1}{s}}-1)^{\frac{1}{2}}}
\end{equation}

where $T_{K}$ is the Kondo temperature.
The fit gives the following values : R$_{0}$ $\approx$ 6.0 $\Omega$, R$_{K0}$ $\approx$ 0.04$\Omega$, $T_{K}$ $\approx$ 19.5 K and $s$ has been fixed at 0.225 which corresponds to the value found in the case of spin 1/2 impurities using numerical renormalization group method \cite{costi94}.

Both Kondo and weak localization scenarios predict a negative magnetoresistance (MR), but the angular dependence is different.
Thus, MR was recorded for different orientations of magnetic field ($B$) and temperatures above and below the temperature
of the resistance minimum (See Fig.5). To avoid complexity due to Lorentz
force anisotropy in the metallic conduction, the applied current is always applied perpendicular to $\vec{B}$. $\theta$ is the
angle between $\vec{B}$ and $\overrightarrow{t}$, the stacking axis,
and $\theta$=0 when $\vec{t}$ and $\vec{B}$ are parallel. For temperature higher
than the temperature of resistance minimum,
the MR is positive and increases with a decreasing temperature, and is almost zero for $\theta$=90 deg. This positive MR is actually consistent with
a metallic Lorentz force effect. As discussed in the supplemental material (III) \cite{supp},
the angular variation is characteristic of a quasi-2D Fermi surface
and can be explained by the small thickness of the SVO layers. In
the Kondo regime, the MR becomes negative ($R\leq R_{0}$), but only
when $\vec{B}$ is almost in the sample plane ($\theta\approx$ 90 deg) (Fig.5).
From the full angular variation of resistivity (see Fig.6a), we show that this
negative MR is restricted to an extremely small angular range. More precisely, after fitting the angular dependence
 with a quasi-2D FS model (Fig.6b), we see that the additional negative MR emerges when the magnetic field is less than 10 degrees from the SL plane. This MR is specially strong
 when the angle between the magnetic field and the plane is less than roughly 5 degrees.
 Note that a negative MR appearing only for the in-plane magnetic fields is opposite to what is expected for a 2D weak localization 
where the orbital effect exists for $\vec{B}$ perpendicular to the carriers trajectory \cite{weakloc}. A negative MR could also arise
 from long-range ordering of the spins, e.g in a ferromagnetic state. However, we find this last possibility very unlikely. This can be inferred from our macroscopic magnetic measurements (see supplemental material ( IV) \cite{supp}), from the DMFT calculations on vanadates heterostructures \cite{millis}, and from the lack of hysteresis in the MR.
Note also that in our geometry with the field perpendicular to the applied current, no anisotropic MR (AMR) of ferromagnetic metals is expected. 

This negative MR is then more likely explained by the dilute Kondo regime that was inferred from the R(T) measurements.
 However, the large anisotropy is rather unexpected since the Kondo MR is isotropic in conventional cases,
such as in many Kondo metals or dilute magnetic conductors \cite{kondoMR,giordanoMR}.
The Kondo MR varies at first order as the square of the (thermally averaged) local magnetization \cite{giordanoMR,alainkondo}.
 If this latter is strongly anisotropic and essentially 2D, then the associated MR should be. 
A similar strongly anisotropic and negative
MR was already reported in LAO/STO which presents
2DEG characteristics \cite{shalom}, and interpreted
in terms of magnetic scattering confined to the interface. Small fluctuating
moments ($\mu\sim$ 0.002 $\mu_{B}$) were also observed by nuclear magnetic
resonance in the LAO/STO interface \cite{salman}, showing the dilute moments consistent with a Kondo scenario.
 A 2DEG created on a single STO doped by a ionic liquid also displays a strongly anisotropic Kondo-like behavior \cite{STOkondo}.
 These different reports in LAO/STO suggest that the large anisotropic Kondo effect is created in a 2DEG close to STO, and that  Ti$^{3+}$  ions carry the magnetic moment \cite{coey,Lee}.
 Here, in our SVO/LVO SLs deposited on STO, the situation is necessarily different since we have shown that there is no evidence of a high mobility 2DEG and that the carriers scattering occurs
in the SVO layers. Consequently, Ti ions can not play a direct role in the anisotropic negative MR. Another explanation is needed to explain the anisotropic Kondo MR observed in vanadate superlattices.

A remarquable quasi-linear dependence of the logarithmic Kondo slope B$_K$ as function of the residual resistivity $\rho_0=\rho_i+\rho_K$  is observed  (inset of Fig.3), where $\rho_K$ is related to the concentration of dilute magnetic impurities.
  In conventional cases, B$_K$ depends only on this "magnetic" residual resistivity $\rho_K$, but we have to discuss if other effects can take place in a SLs. 
Indeed, Giordano \textit{et al.} \cite{giordano2} have demonstrated that the Kondo in small systems
effect can be different from its bulk counterpart, and can depend also on the level of non-magnetic disorder, e.g $\rho_i$.
 Specifically, the Kondo effect is quenched in the limit of large disorder (large residual resistivity) \cite{giordano2}. The anisotropy
of the moments distribution can be also reinforced by the effect of
multiple scattering of conduction electrons from both local moments
and the interface, which results in an effective uniaxial out-of-plane
anisotropy \cite{ujsaghy}. This approach can explain both that B$_K$ is sensitive to all the disorder quantified in $\rho_0$ and that the Kondo MR is anisotropic,
 but the predicted variations are quantitatively opposite to our results.
We observe a largest Kondo effect for the largest disorder, and the negative MR appears only for in-plane magnetic fields, which further confirms that our results can not
be explained by the size and disorder corrections to the Kondo effect. Thus, a bulk-like interpretation which deals with magnetic impurities only seems more appropriate. 
 In this framework, the quasi-linear variation of B versus $\rho_0$  implies that almost all impurities
 tuned by the growth temperature contribute to $\rho_K$, and create a magnetic scattering center. For its part, the ordinary impurities contribution is almost constant and is $\rho_i $ $\approx$ 95 $\mu\Omega$.cm as seen in the inset of Fig.3. 
 
The possible origins of the magnetic impurities can be discussed from our recent spectroscopy analysis.
 In fact, we have observed that the mean valence
 state of vanadium varies over 3 unit cells at the interface between LVO and SVO layers \cite{tan13}. 
It means that mixed valency V$^{3+}$ and V$^{4+}$ are present in the same layer. 
As a consequence, the vanadium in the lower
oxydation state can be stabilized by oxygen vacancies in the SrVO$_3$ layers, with one electron staying localized close to the vacancy, 
and playing the role of a spin 1/2  Kondo center. As discussed in the supplemental material (I) \cite{supp}, the expansion of the out-of-plane lattice parameter for the SL deposited at low temperature is consistent with a larger oxygen vacancies concentration that forms in the layers during the ablation process.
This could also explain the largest $\rho_0$ and the strongest Kondo effect: when the growth temperature increases, oxygen vacancies tend to diffuse in the substrate due to the affinity of STO to absorb them \cite{kalabukov, nous}, and finally the number of Kondo centers present in the SVO layers strongly decreases. 

To conclude, we have studied the magnetotransport (MR) properties
of a series of (LaVO$_{3}$)$_{6}$/(SrVO$_{3}$)$_{6}$ superlattices. We show that the electronic conduction and galvanomagnetic properties are due 
to the  metallic SrVO$_{3}$ layers with a quasi-2D Fermi surface.
At low temperature, a clear signature of Kondo scattering is observed with a huge anisotropic negative magnetoresistance (MR) similar to reports
 at the LaAlO$_3$/SrTiO$_3$ interface. 
We explain this effect by the valency changes of V$^{3+}$ for some V$^{4+}$
in the SrVO$_{3}$ layer which partially compensate the oxygen vacancies created during the film growth.
 The observation of such large anisotropic MR in a system other than the LaAlO$_3$/SrTiO$_3$ interface
 shows that the existence of a 2DEG and the presence of the titanium ion are not mandatory for this property.
 It can be observed with another \emph{d} magnetic ion with mixed valency such as Vanadium and if oxygen vacancies are present.

We acknowledge partial support of the French Agence Nationale de la Recherche
(ANR) through the program Investissements d'Avenir (ANR-10-LABX-09-01) and the LABEX EMC3.
The authors thank F. Veillon for technical support during the transport
measurement and J. Lecourt for the target preparation for the PLD.

\newpage 

\begin{figure}
\includegraphics[width=8cm]{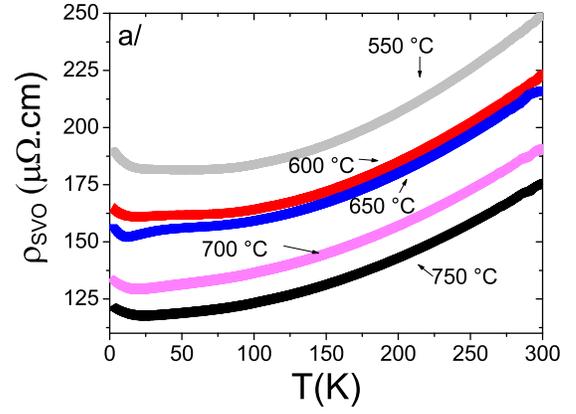}
\includegraphics[width=8cm]{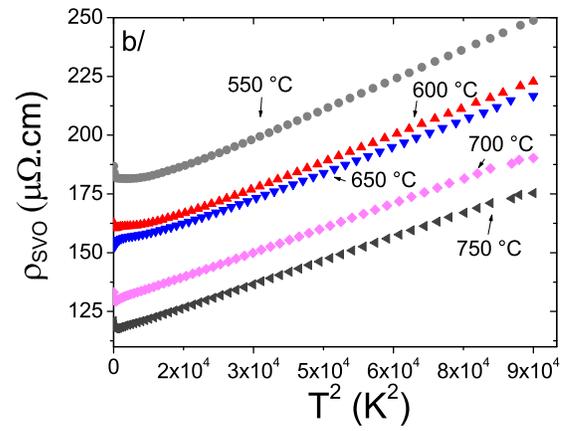}
\caption{a/ Resistivity of the SVO layers as function of temperature for the superlattice series
 grown at different temperatures. From top to bottom, T$_{growth}$= 550, 600, 650, 700, 750$^{\circ}$ C. b/ Resistivity of the SVO layers as function of the temperature square.}
\end{figure}

\begin{figure}
\includegraphics[width=8cm]{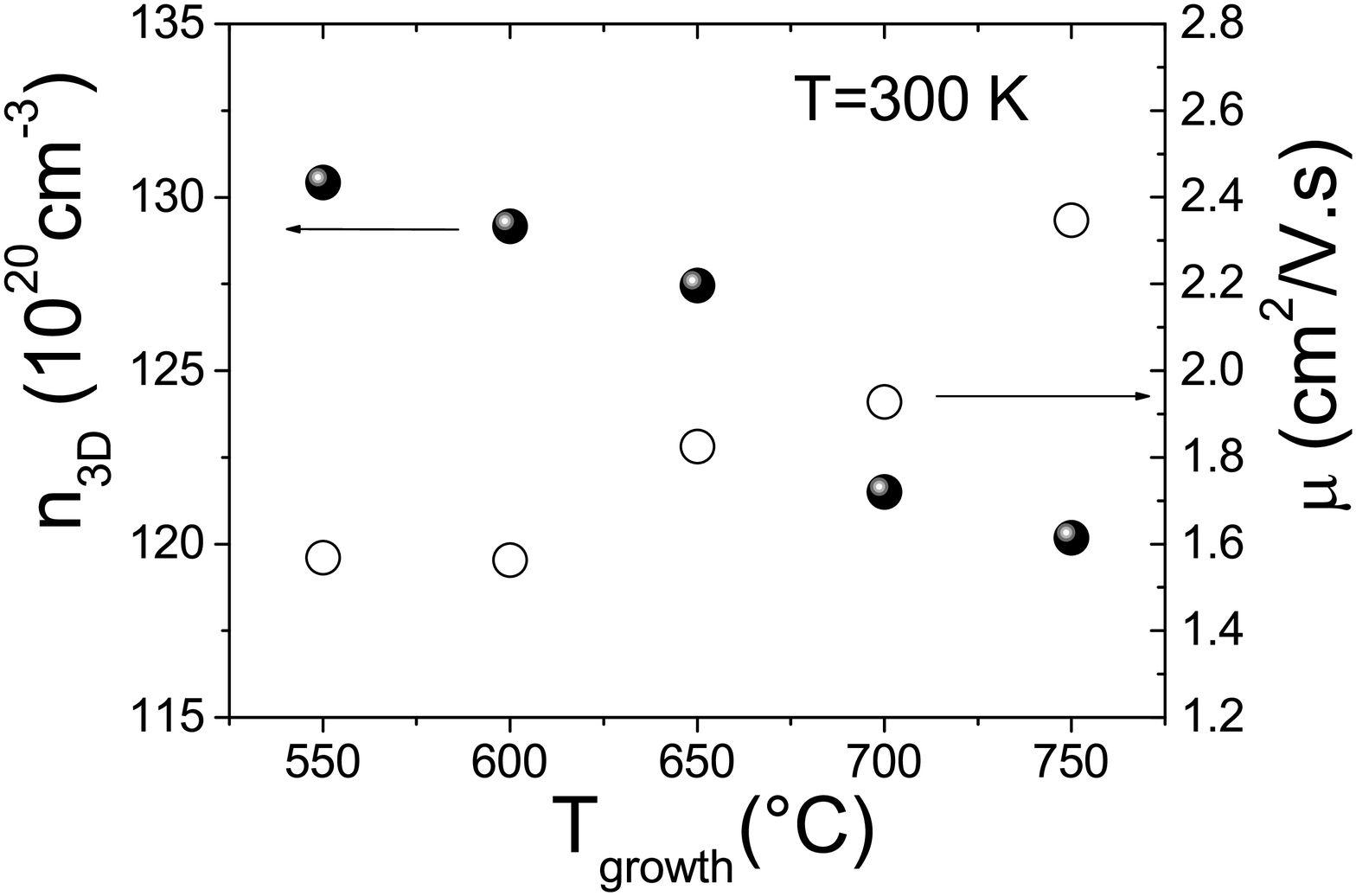}
\caption{Carriers density and electronic mobility as function of growth temperature for the LVO/SVO SLs. Data are measured at $T=$300 K.}
\end{figure}

\begin{figure}
\centering\includegraphics[width=8cm]{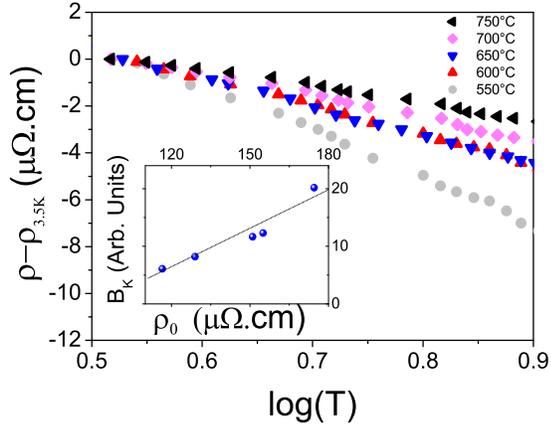}
\caption{Low temperature resistivity as function of the logarithmic of the temperature for
 the SL series, showing a -B$_K$.log(T) form in the intermediate temperature range. From bottom to top,
 T$_{growth}$=550, 600, 650, 700, 750 $^{\circ}$ C. The inset depicts the slope of B$_K$ as function of the residual resistivity, $\rho_{0}$.} 
\end{figure}

\begin{figure}
\centering\includegraphics[width=8cm]{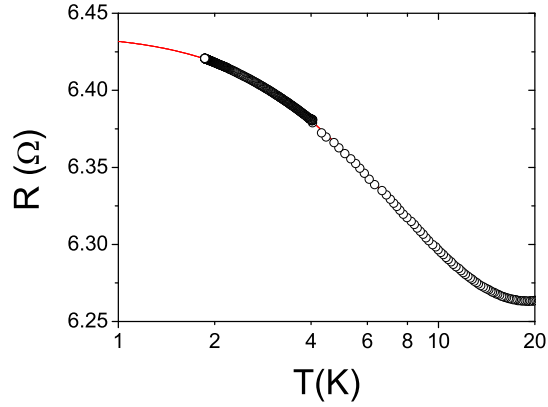}
\caption{Low temperature part of the resistivity as function of the logarithmic of the temperature for
 a superlattice grown at 750$^{\circ}$ C. The solid line is a fit using a Kondo contribution for spin 1/2 impurities (equations (1)-(3)).}
\end{figure}

\begin{figure}
\centering\includegraphics[width=8cm]{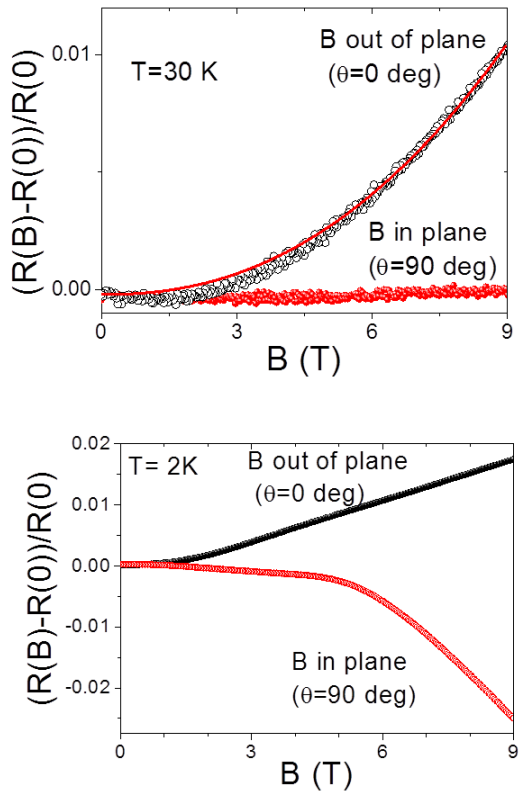}
\caption{Magnetoresistance MR as function of the magnetic field B for two directions. Top (T=30 K) : A classical (the plain  curve is a typical B$^2$ variation)
 but anisotropic MR is observed as for a quasi 2D Fermi surface (see supp. mat). Bottom (T=2 K) : A negative MR appears for B in the plane of the superlattice.}
\end{figure}

\begin{figure}
\centering\includegraphics[width=8cm]{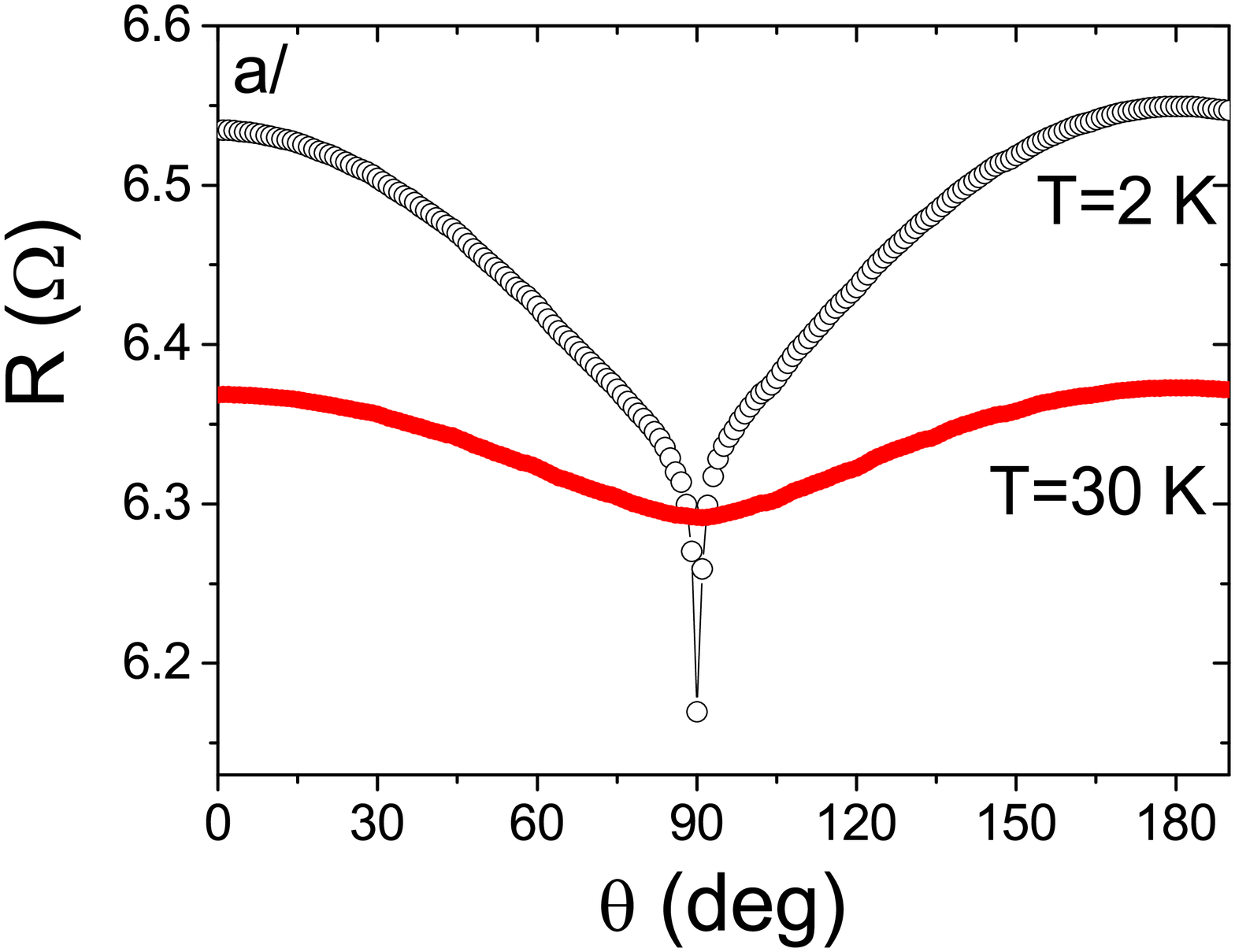}
\centering\includegraphics[width=8cm]{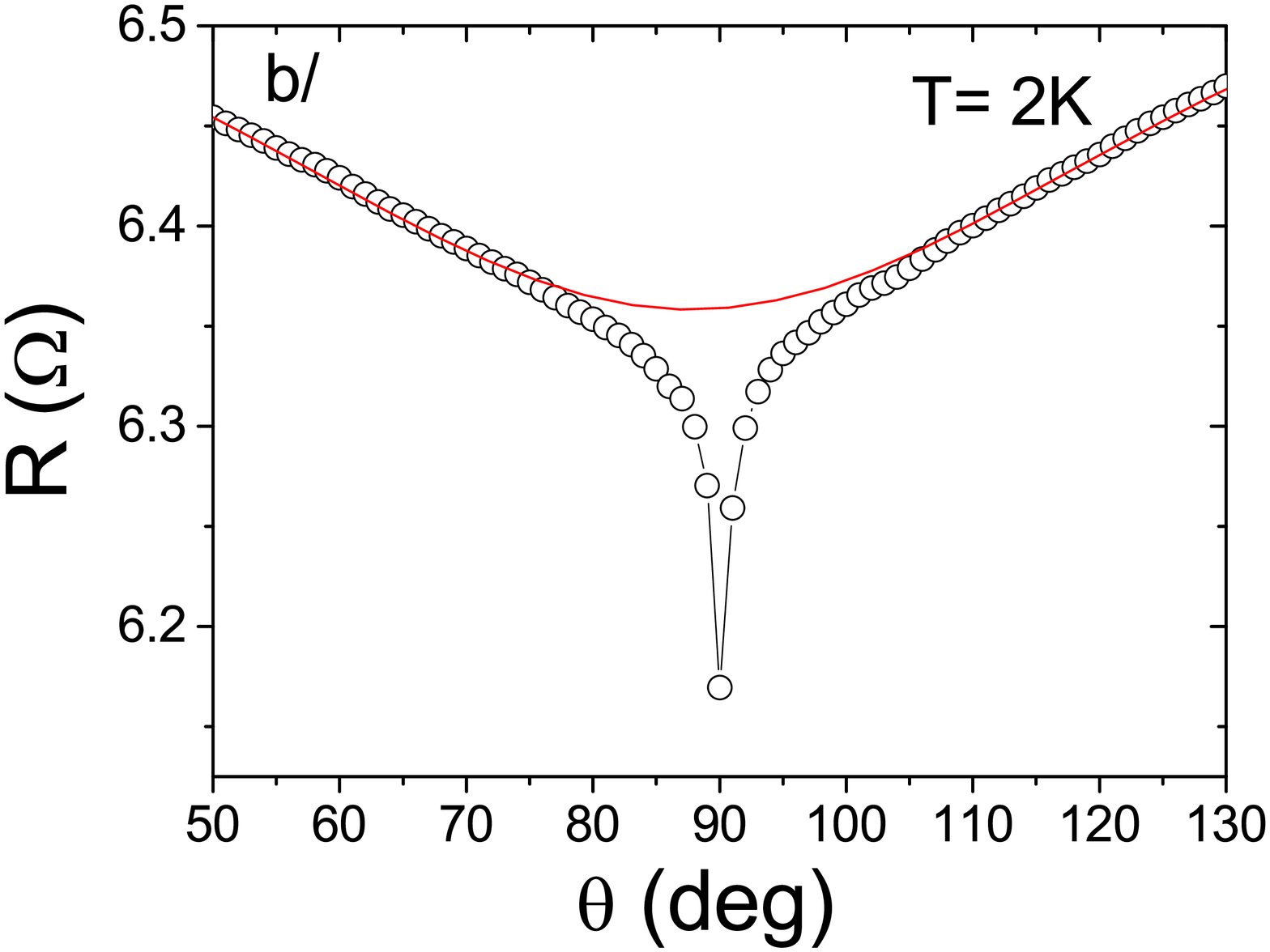}
\caption{a/ Angular variation of resistivity under a 9T magnetic field for a superlattice grown at 750$^{\circ C}$ (T=30 K and 2 K).
 Note the sharp minimum corresponding to the negative magnetoresistance at low temperature for $\theta$ close to 90 deg. b/ A zoom on the 2K variation. The solid line is a fit with corresponding to a quasi 2D Fermi Surface (see supp. mat III \cite{supp}). The negative MR appears within a $\pm$ 10 deg range, and is specially marked within a $\pm$ 5 deg range.}
\end{figure}


\begin{thebibliography}{10}


\bibitem{ohtomo04} A.~Ohtomo, H.~Y. Hwang, Nature 427, 423 (2004).

\bibitem{nakagawa06} N.~Nakagawa, H.~Y. Hwang, D.~A. Muller, Nature
Mater. 5, 204 (2006).

\bibitem{reyren07} N.~Reyren, S.~Thiel, A.~D. Caviglia, L.~F.
Kourkoutis, G.~Hammerl, C.~Richter, C.~W. Schneider, T.~Kopp,
A.-S. Ruetschi, D.~Jaccard, M.~Gabay, D.~A. Muller, J.-M. Triscone,
J.~Mannhart, Science 317, 1196 (2007).

\bibitem{kourkoutis06} L.~Fitting~Kourkoutis, Y.~Hotta, T.~Susaki,
H.~Y. Hwang, D.~A. Muller, Phys. Rev. Lett. 97, 256803 (2006).

\bibitem{kalabukov} A. Kalabukhov, R. Gunnarsson, J. Borjesson, E. Olsson, T. Claeson, and D. Winkler, Phys. Rev. B 75, 121404 (2007).

\bibitem{ariando1} Ariando, X. Wang, G. Baskaran, Z. Q. Liu, J. Huijben, J. B. Yi, A. Annadi, A. Roy Barman, A. Rusydi, S. Dhar, Y. P. Feng, J. Ding, H. Hilgenkamp and T. Venkatesan,  Nature Communications  2, 188 (2011).

\bibitem{bert} J.A. Bert, B.Kalisky, C.Bell, M. Kim, Y. Hikita, H.Y. Hwang, K.A. Moler, Nat. Phys. 7, 767 (2011).

\bibitem{brinkman} A. Brinkman, M. Huijben, M. van Zalk, J. Huijben, U. Zeitler, J.C. Maan, W.G. van der Wiel, G. Rijnders, D.H.A. Blank, H. Hilgenkamp,  Nature Materials 6, 493 (2007).

\bibitem{shalom}M. Ben Shalom, C. W. Tai, Y. Lereah, M. Sachs, E. Levy, D. Rakhmilevitch, A. Palevski, and Y. Dagan,  Phys. Rev. B 80, 140403(R) (2009).

\bibitem{annealing} M. Salluzzo, S. Gariglio, D. Stornaiuolo, V. Sessi, S. Rusponi, C. Piamonteze, G. M. De Luca, M. Minola, D. Marre, A. Gadaleta, H. Brune, F. Nolting, N. B. Brookes, and G. Ghiringhelli, Phys. Rev. Lett. 111, 087204 (2013).

\bibitem{coey}  J.M.D. Coey, Ariando and W.E. Pickett. Magnetism at the edge: New phenomena at oxide interfaces. MRS Bulletin 38,  1040 (2013). 

\bibitem{inaba95} F.~Inaba, T.~Arima, T.~Ishikawa, T.~Katsufuji, Y. Tokura, Phys. Rev. B 52,  R2221  (1995).

\bibitem{miyasaka00} S.~Miyasaka, T.~Okuda, Y.~Tokura, Phys. Rev.Lett. 85, 5388 (2000).

\bibitem{sayer75} M.~Sayer, P.~Chen, R.~Fletcher, A.~Mansingh, J. Physics C 8, 2059 (1975).

\bibitem{sheets07} W.~C. Sheets, B.~Mercey, W.~Prellier, Appl. Phys. Lett. 91, 192102 (2007).

\bibitem{boullay11} P.~Boullay, A.~David, W.~C. Sheets, U.~Luders, W.~Prellier, H.~Tan, J.~Verbeeck, G.~V. Tendeloo, C.~Gatel, G.~Vincze, Z.~Radi, Phys. Rev. B 83, 125403 (2011).

\bibitem{tan13} H.~Tan, R.~Egoavil, A.~Beche, J.~Verbeeck, G.~V. Tendeloo, H.~Rotella, P.~Boullay, A.~Pautrat, W.~Prellier, Phys. Rev. B 88, 155123 (2013).

\bibitem{supp} See Supplemental Material for details on structural characterisations, and on analysis of transport and magnetic properties of the superlattices.

\bibitem{nunez} G. Garbarino and M. Nunez-Regueiro, Solid State Commun. 142, 306 (2007).

\bibitem{alain} A. Pautrat and W. Kobayashi,  EPL 97,  67003 (2012).

\bibitem{reyes00} D.~R. Ardila, J.~Andreeta, H.~Basso, J. Crystal Growth 211, 313 (2000).

\bibitem{gu} M. Gu, S. A. Wolf, and J. Lu,  Adv. Mater. Interfaces 1, 1300126 (2014)

\bibitem{Yoshi} K. Yoshimatsu, T. Okabe, H. Kumigashira, S. Okamoto,
S. Aizaki, A. Fujimori, and M. Oshima, Phys. Rev. Lett. 104, 147601
(2010).

\bibitem{LVOMIT} Zhicheng Zhong, Markus Wallerberger, Jan M. Tomczak, Ciro Taranto, Nicolaus Parragh, Alessandro Toschi, Giorgio Sangiovanni, Karsten Held, arXiv:1312.5989 (2013).

\bibitem{costi94} T.~A. Costi, A.~C. Hewson, J. Phys. Cond. Matter. 6, 2519 (1994).

\bibitem{weakloc} P.A. Lee and T.V. Ramakrishnan, Rev. Mod. Phys. 5, 287 (1985).

\bibitem{millis} Hung T. Dang and Andrew J. Millis, Phys. Rev. B 87, 184434 (2013).

\bibitem{kondoMR} A.K.Nigam and A.K.Majumdar, Phys.Rev.B 27, 495 (1983), and references cited therein.

\bibitem{giordanoMR} N. Giordano, Phys. Rev. B 53, 2487 (1996).

\bibitem{alainkondo} A. Pautrat and W. Kobayashi, Phys. Rev. B 82, 115113 (2010).

\bibitem{salman} Z. Salman, O. Ofer, M. Radovic, H. Hao, M. Ben Shalom,
K.H. Chow, Y. Dagan, M.D. Hossain, C.D.P. Levy, W.A. MacFarlane, G.M.
Morris, L. Patthey, M.R. Pearson, H. Saadaoui, T. Schmitt, D. Wang,
and R.F. Kiefl, Phys. Rev. Lett. 109, 257207 (2012).

\bibitem{STOkondo} Menyoung Lee, J.R. Williams, Sipei Zhang,C. Daniel Frisbie and D. Goldhaber-Gordon, Phys. Rev. Lett. 107,  256601 (2011).

\bibitem{Lee} J.-S Lee et al., Nature Materials 12, 703 (2013).

\bibitem{giordano2} M. A. Blachly and N. Giordano, Phys. Rev. B 51, 12537 (1995).

\bibitem{ujsaghy} O. Ujsaghy and A. Zawadowski, Phys. Rev. B 57, 11598 (1998).

\bibitem{nous} H. Rotella, O. Copie, A. Pautrat, P. Boullay, A. David, D. Pelloquin, C. Labbe, C. Frilay and W. Prellier,  J. Phys.: Condens. Matter 27, 095603 (2015).

\end{thebibliography}
\end{document}